\documentclass[aps,prb,twocolumn,showpacs,superscriptaddress]{revtex4}
\usepackage{graphicx}
\usepackage{amssymb}
\usepackage{epstopdf}
\usepackage{amsmath}
\begin{document}

\preprint{APS/123-QED}

\title{Magetic and Superconducting Properties of Single Crystals of Fe$_{1+\delta}$Te$_{1-x}$Se$_{x}$ System}
\author{Jinhu Yang} \email{yangjinhu@kuchem.kyoto-u.ac.jp}
\affiliation{Department of chemistry, Graduate  School of Science, Kyoto University, 606 - 8502, Japan }
\author{Mami Matsui}
\affiliation{Department of chemistry, Graduate  School of Science, Kyoto University, 606 - 8502, Japan }
\author{Masatomo Kawa}
\affiliation{Department of chemistry, Graduate  School of Science, Kyoto University, 606 - 8502, Japan }
\author{Hiroto Ohta}
\affiliation{Department of chemistry, Graduate  School of Science, Kyoto University, 606 - 8502, Japan }
\author{Chishiro Michioka}
\affiliation{Department of chemistry, Graduate  School of Science, Kyoto University, 606 - 8502, Japan }
\author{Chiheng Dong }
\affiliation{Department of physics, Graduate  School of Science, Zhejiang University, Hangzhou 310027,China} 
\author{Hangdong Wang}
\affiliation{Department of physics, Graduate  School of Science, Zhejiang University, Hangzhou 310027,China} 
\author{Huiqiu Yuan}
\affiliation{Department of physics, Graduate  School of Science, Zhejiang University, Hangzhou 310027,China} 
\author{ Minghu Fang} 
\affiliation{Department of chemistry, Graduate  School of Science, Kyoto University, 606 - 8502, Japan }
\affiliation{Department of physics, Graduate  School of Science, Zhejiang University, Hangzhou 310027,China} 
\author{Kazuyoshi Yoshimura}\email{kyhv@kuchem.kyoto-u.ac.jp}
\affiliation{Department of chemistry, Graduate  School of Science, Kyoto University, 606 - 8502, Japan }






\date{\today}

\begin{abstract}
The spin-fluctuation effect in the Se-substituted system Fe$_{1+\delta}$Te$_{1-x}$Se$_x$ ($x$ = 0, 0.05, 0.12, 0.20, 0.28, 0.33, 0.45, 0.48 and 1.00; $0<\delta < 0.12$) has been studied by the measurements of the X-ray diffraction, the magnetic susceptibility under high magnetic fields and the electrical resistivity under magnetic fields up to 14 T. The samples with $x$ = 0.05, 0.12, 0.20, 0.28, 0.33, 0.45 and 0.48 show superconducting transition temperatures in the ranger of 10 K$\sim$14 K. We obtained their intrinsic susceptibilities by the Honda-Owen method. A nearly linear-in-$T$ behavior in magnetic susceptibility of superconducting samples was observed, indicating the antiferromagnetic spin fluctuations have a strong link with the superconductivity in this series. The upper critical field $\mu_0H_{c2}^{orb}$ for $T\to$ 0 was estimated to exceed the Pauli paramagnetic limit. The Kadowaki-Woods and Wilson ratios indicate that electrons are strongly correlated in this system. Furthermore, the superconducting coherence length and the electron mean free path were also discussed. These superconducting parameters indicate that the superconductivity in the Fe$_{1+\delta}$Te$_{1-x}$Se$_x$ system is unconventional.
\end{abstract}



\maketitle

\section{INTRODUCTION}
Shortly after the discovery of the iron-based oxypnictide superconductor LaFeAsO$_{1-x}$F$_x$ with $T_c$ of 26 K, \cite{Hosono} another family of the iron-based chalcogenide superconductor $\alpha$-FeSe with $T_c$ of about 8.5 K was reported by Wu's group. \cite{Wu} Later on, Fang's group enhanced the $T_c$ up to 14 K by substituting Se for Te in the FeTe$_{1-x}$Se$_x$ system. \cite {Fang} Interestingly, although both FeTe and FeS are not superconductors under ambient pressure, the superconductivity can be induced by Se doping in FeTe$_{1-x}$Se$_x$ as well as S doping in the FeTe$_{1-x}$S$_{x}$ system. \cite{Fang, STe, Mizuguchi} While there is no sign of superconductivity in pure FeTe under high pressure in contrast to that the superconducting transition temperature was enhanced to 37 K, just below the McMillan limit in FeSe under high pressure. \cite{37K} Superconducting $\alpha$-Fe$_{1.01}$Se belongs to the tetragonal symmetry system at room temperature and undergoes a structural transition to an orthorhombic phase at 90 K, while the non-superconducting Fe$_{1.03}$Se does not. \cite{McQueen} FeSe has a simpler structure by stacking only the conducting Fe$_2$Se$_2$ layers in contrast to the Fe-As based superconductors having both the conducting Fe$_2$As$_2$ layers and the blocking R$_2$O$_2$ layers (R = rare earth elements). FeTe has the same structure as FeSe but with a rather complex magnetic structure. For example, the stoichiometric sample FeTe has an commensurate antiferromagetic (AF) ordering at low temperatures after suffering a structural phase transition, while samples with excess iron Fe$_{1+\delta}$Te has an incommensurate AF ordering. \cite{Bao} According to the result of density functional calculations, the spin density wave (SDW) is more stable in FeTe than that in FeSe. \cite{Subedi} Therefore, the doped sample FeTe is expected to have higher $T_c$. This was indeed observed in S or Te substituted systems for Se in FeSe. \cite{SSe} The iso-valent substitution does not directly introduce extra carriers but may change the topology of the Fermi surface. \cite{Fang} Recently, a spin fluctuation spectrum and a spin gap behavior were observed by neutron scattering. \cite{Bao, nuclear} In fact, the NMR results indicated that the AF spin fluctuations were enhanced greatly toward $T_c$, indicating the importance of AF spin fluctuations for the superconducting mechanism in FeSe. \cite{Cava}

As for the sample preparations, the iron-based superconductors, as previously reported, however, usually contain a very small amount of magnetic impurities, e.g., Fe$_7$Se$_8$ and Fe$_3$O$_4$. \cite{Wu,Fang, Li, Mizuguchi, McQueen, Kazumasa, Williams, Taen} Therefore, in many cases the Verwey-phase like transition happens around 120 K due to the existence of the magnetic impurity Fe$_3$O$_4$ which causes a peak in the magnetic susceptibility. \cite{Fe3O4} To elucidate the superconducting mechanism and its relation with the AF spin fluctuations, it is vitally important to obtain the intrinsic susceptibility. In this study, we successfully synthesized the single crystals of Fe$_{1+\delta}$Te$_{1-x}$Se$_{x}$ and measured their magnetic susceptibilities under high magnetic fields and obtained the intrinsic susceptibilities by using the Honda-Owen plot. \cite{Honda} We have found that the magnetic susceptibility of Fe$_{1+\delta}$Te$_{1-x}$Se$_x$ ($0.12 \leqslant x \leqslant 1.00$) decreases with decreasing temperature from 300 K to 20 K, similar to the results in high temperature cuprate superconductor (La$_{1-x}$Sr$_x$)$_2$CuO$_4$, or Fe-As based superconductors. \cite{Takagi, Klingeler} In addition, we conducted electrical resistivity measurements of two single crystal samples with very close composition $x\sim$ 0.3, both of which show the superconductivity at $T_c$ $\sim$ 14 K under magnetic fields up to 14 T in order to estimate the upper critical field $\mu_0H_{c2}^{orb}$ and the coherence length $\xi$. As a result, the upper critical field is found to be much larger than the Pauli limit, and the initial slope near $T_c$ is comparable with those of Fe-As based superconductors. \cite{Kohama, Terashima} Although Fe(Te-Se) is a layered superconductor, both the upper critical field and the initial slope near $T_c$ show weak anisotropies. In order to investigate the electron correlation strength, we have estimated Kadowaki-woods and Wilson ratios which indicate a strongly correlated electrons picture. The superconducting coherence length and the electron mean free path are also discussed, leading to the fact that Fe$_{1+\delta}$Te$_{1-x}$Se$_{x}$ is a clean superconductor.
\section{EXPERIMENTAL}
The high-quality single crystals of Fe$_{1+\delta}$Te$_{1-x}$Se$_{x}$ ( $x$ = 0, 0.05, 0.12, 0.20, 0.28, 0.33, 0.45, 0.48 and 1.00; $0<\delta<0.12$) were prepared from Fe powders (4N purity), Te powders (4N purity) and Se powders (5N purity). Stoichiometric quantities of about 3g-mixtures were loaded into a small quartz tube. This small tube was then sealed into second evacuated quartz tube, and placed in a furnace at room temperature. The temperature was slowly ramped up to 920$^{\circ}$C over for 36 hours and then held at that temperature for another 12 hours in order to obtain sample homogeneities. Then, the temperature was reduced to 400$^{\circ}$C over for140 hours. On the other hand, the polycrystalline of Fe$_{1+\delta}$Se was synthesized by previously reported solid state reaction method. \cite{Wu}  
 
The obtained single crystal samples were ground into powders for measuring powder X-ray diffraction (XRD) with Cu $K_\alpha$ radiation. The detailed structural parameters were analyzed by Rietveld refinements. The compositions of the single crystals were analyzed using SEM (JED-2300, JEOL) equipped with an Energy Dispersive X-Ray (EDX) spectrometer. The DC magnetic measurements were performed by using a Superconducting Quantum Interference Device (SQUID, Quantum Design Magnetometer). For the observations of the superconducting transitions, both the zero-field cooling (ZFC) and field cooling (FC ) measurements were performed under the magnetic field of 20 Oe. The temperature dependence of resistivity was measured using a standard dc four-probe method under dc magnetic fields up to 14 T with a Physics Property Measurements System (PPMS, Quantum Design Magnetometer). The current direction was parallel to the a- axis of the single crystal sample. 

 \section{RESULTS AND DISCUSSION}
  \subsection{XRD and EDX Spectroscopy}
 The obtained single crystal has the layered planes held together by Van der Waals force only, and thus the crystal can easily be cleaved. X-ray diffraction pattern of a typical single crystal Fe$_{1.12}$Te$_{0.72}$Se$_{0.28}$ measured with the scattering vector perpendicular to the cleaved surface was shown in Fig. 1 (a) and the image of the single crystal is in the inset. Only (00$l$) reflections appear, indicating that the c-axis is perpendicular to the cleaved surface. In order to get more structural information from the XRD pattern, the singe crystal were ground into powders for powder XRD measurements. Figure 1 (b) shows XRD patterns of the selected samples of Fe$_{1+\delta}$Te$_{1-x}$Se$_x$( $x$ = 0, 0.05, 0.12, 0.20 and 0.33). The real compositions were analyzed by EDX measurements as listed in Table I. The EDX spectroscopy results show that there is a slight excess amount of Fe existing in each sample, and furthermore, the Se content has a smaller ratio than the nominal one. All the peaks are well indexed based on a tetragonal cell with the space group of $P$4/$nmm$, except for a small amount of the magnetic impurity phase of Fe$_7$Se$_8$-type, indicating that the samples are almost in a single phase. The impurity phase exits in the surface or in the inter-layers of the single crystal. It should be pointed out that there is a very small amount of Fe$_3$O$_4$ impurity in each sample, which cannot however be probed in the XRD patterns but causes a peak in temperature dependence of magnetic susceptibility at about 120 K due to the Verwey phase transition. \cite{Fe3O4} With increasing Se doping level, the a-axis decreases slightly, while the c-axis shrinks remarkably, which makes the (001) and (200) diffraction peaks shift to higher angles monotonously, shown as enlarged views in the inset. The cell volume is consequently decreased by substituting Te for Se from 92 \AA$^3$ to 78 \AA$^3$, indicating that the samples are in a solid solution in which Se enters the lattice as Fe$_{1+\delta}$Te$_{1-x}$Se$_x$ successfully.   
  \begin{figure}[htbp] 
   \centering
    \includegraphics[width= 8cm]{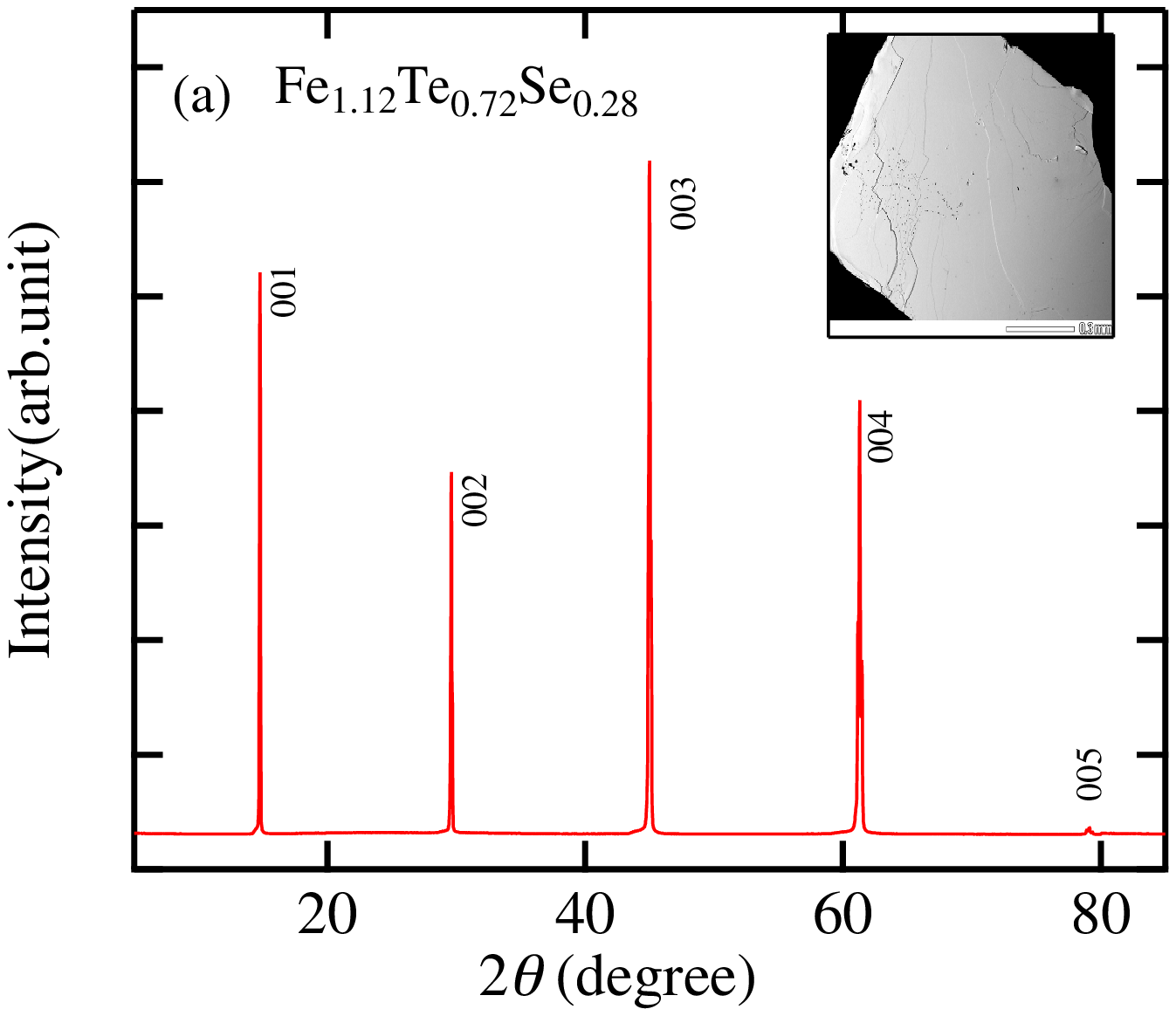} 
    \includegraphics[width= 8cm]{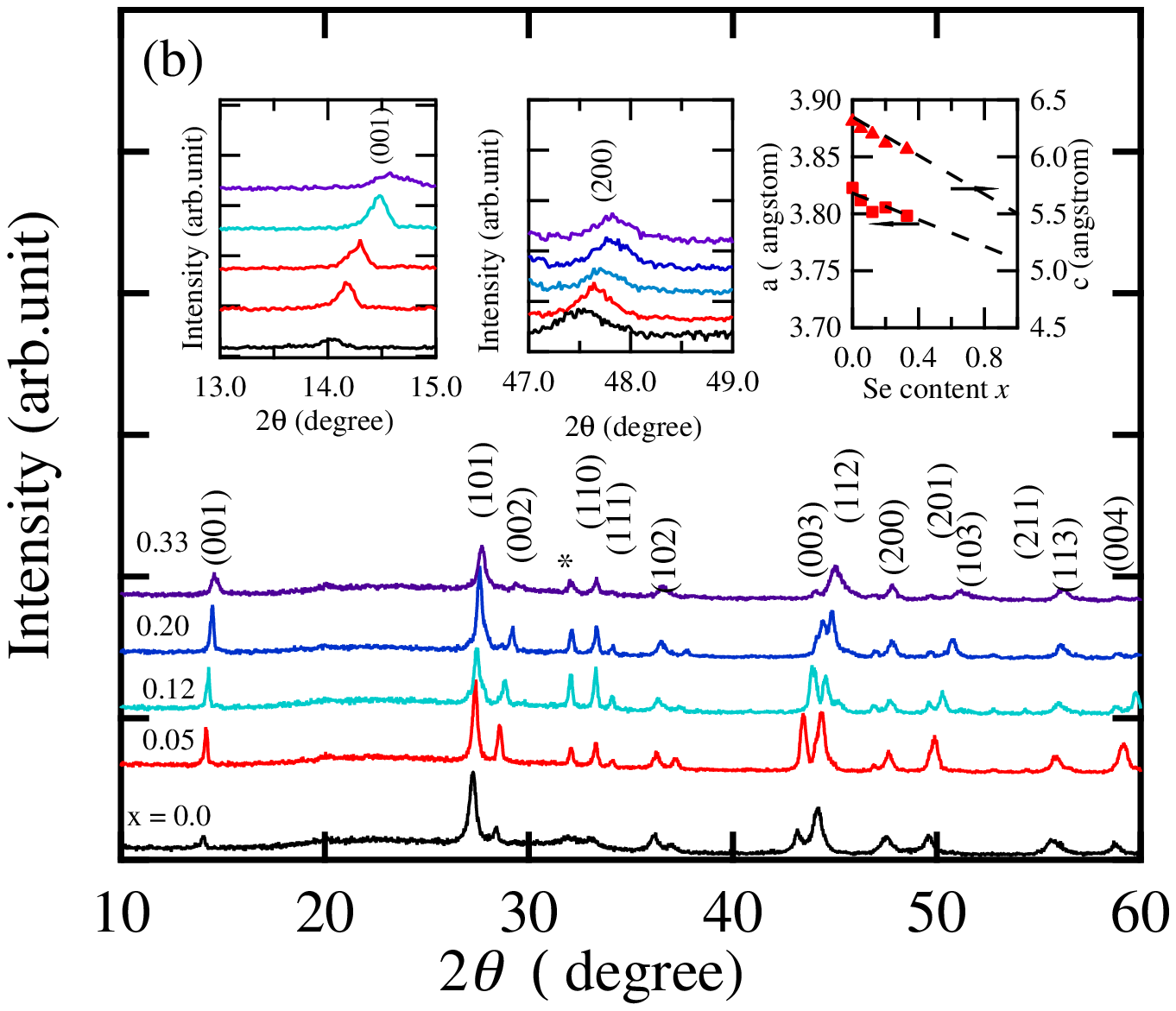} 
    \caption{X-ray diffraction patterns of Fe$_{1+\delta}$Te$_{1-x}$Se$_{x}$. (a) A typical single crystal XRD pattern for sample Fe$_{1.12}$Te$_{0.72}$Se$_{0.28}$ as well as the single crystal image in the inset. (b) Powder XRD patterns by using samples of ground single crystal; the peaks marked by * are Fe$_7$Se$_8$ impurity phase. The enlarged view of the (001) and (200) peaks and the lattice constants as functions of Se content $x$ were shown in the inset. }
  
\end{figure}
\begin{figure}[htbp]    \centering
   \includegraphics[width= 8cm]{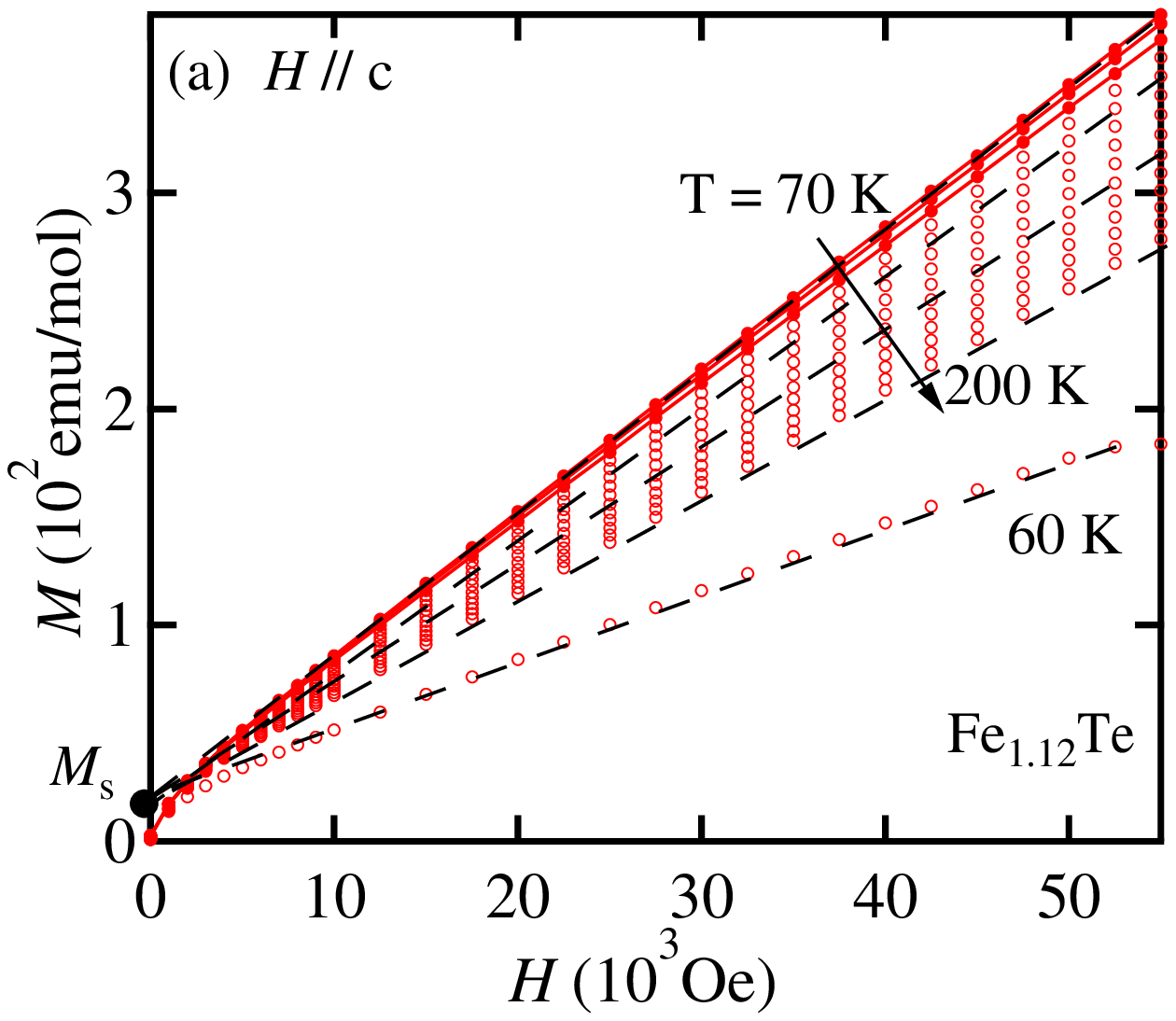} 
   \includegraphics[width= 8cm]{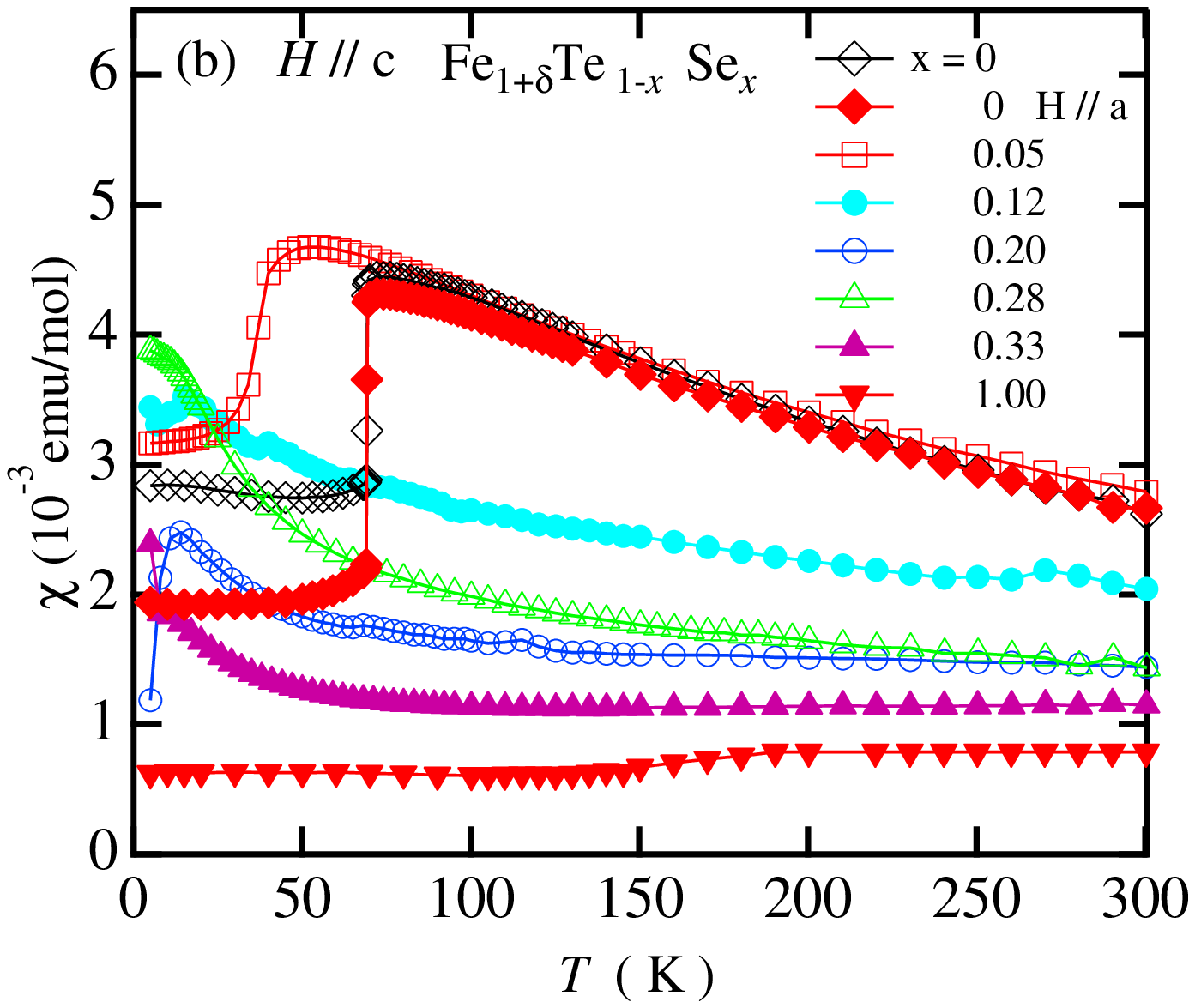}
   \includegraphics[width= 8cm]{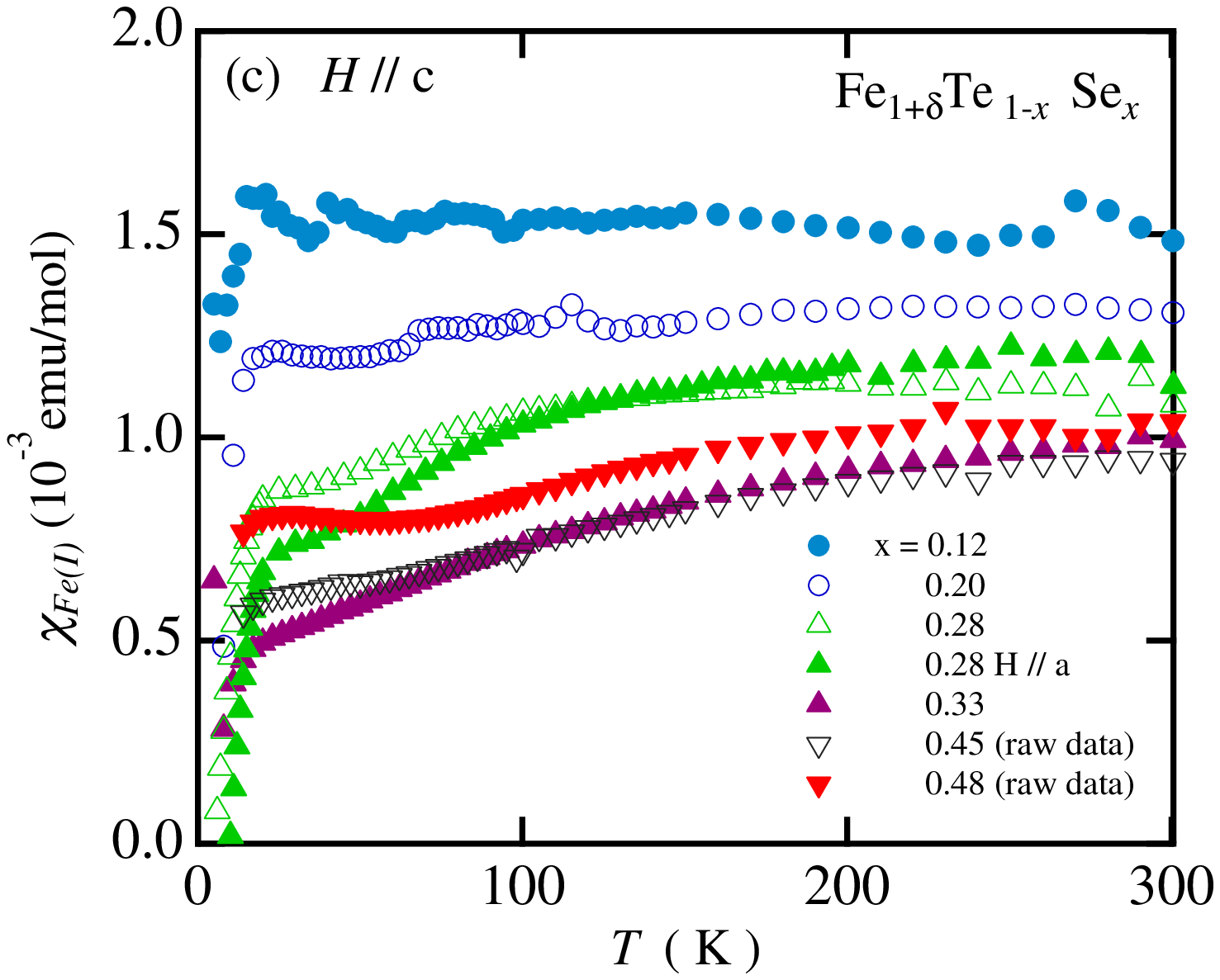}
  \label{fig:1}
  \caption{(a) Isothermal magnetization ($M$) with magnetic field ($H$) in the temperature range 60 K $\leqslant T \leqslant$ 200 K with the step of 10 K, $M_s$ is saturation moment of the impurities. (b) Temperature dependence of the intrinsic susceptibility for Fe$_{1+\delta}$Te$_{1-x}$Se$_x$ ($x$ = 0, 0.05, 0.12, 0.20, 0.28 and 0.33) with $H$//c. The intrinsic susceptibility of Fe$_{1.12}$Te was also measured under $H$//a. (c) The Fe(I) site contribution to the magnetic susceptibility: $\chi_{Fe(I)}$ as a function of temperature for superconducting samples with $x$ = 0.12, 0.20, 0.28, 0.33, 0.45 and 0.48 with $H$//c as well as the sample with $x$ = 0.28 with $H$//a. The anisotropy susceptibility in this series is very weak: the susceptibility in sample of $x$ = 0.28 showed only weak dependence on the magnetic direction of $H$//c or $H$//a.}
\end{figure}
\subsection{Magnetic Susceptibility}
Since the presence of a small amount of ferromagnetic impurities which have a profound effect on the low-field magnetization as shown in Fig. 2 (a) for sample Fe$_{1+\delta}$Te. A linear-in-$H$ term in high magnetic fields magnetization appears in each curve for various temperatures; if we extrapolate the data from high magnetic fields to $H$ = 0, all the extrapolation ends into almost the same point, $M_s$, indicating the saturation magnitude moment of the impurity. According to the Honda-Owen plot, by extrapolating the measured susceptibility $M/H$ = $\chi$ + $C_s$$M_s/$$H$ for $1/H$ $\to$ 0, where $M/H$ is the measured susceptibility, $\chi$ the intrinsic susceptibility, $C_s$ the presumed ferromagnetic impurity content and $M_s$ its saturation magnetization. The influence of ferromagnetic impurities must be avoided in order to obtain the intrinsic susceptibility, $\chi$. Therefore, we use the Honda-Owen method to obtain the $\chi$ of these samples as $\chi(T)$ = $ \Delta M \over \Delta H$. The magnetizations were measured between 5 K and 300 K separately under magnetic fields of 3 and 4 T, or 4 and 5 T, above which the magnetizations of the magnetic impurity were supposed to be saturated. Figure 2 (b) shows the temperature dependence of the intrinsic magnetic susceptibility for samples with $x$ = 0, 0.05, 0.12, 0.20, 0.28, 0.33 and 1.00. The external field is applied parallel along the c-axis. For undoped sample Fe$_{1.12}$Te, the magnetic susceptibility $\chi(T)$ increases with decreasing temperature, and decreases sharply near 69 K, due to the AF phase transition accompanied by the structural phase transition, then becomes almost the constant with decreasing temperature, in agreement with the previous reports. \cite{Wang} The susceptibility does not show any anisotropy since it has almost no distinct difference in the paramagnetic phase in cases of $H$//c and $H$//a, and even at low temperatures $\chi^{H//c}$(5 K)/$\chi^{H//a}$(5 K) $\sim$ 1.45 shows a weak anisotropy. For the superconducting sample, the intrinsic susceptibility of the sample with $x$ = 0.28 under $H$//c and $H$//a also shows a very weak anisotropy as displayed in Fig. 2 (c). On Se-doping, the AF transition shifts to lower temperature and the peak is broadened, then is hardly observed in the range $x>$ 0.12, where the superconducting phase transition occurs at 10 K $\sim$ 14 K. The upturn of $\chi(T)$ at low temperatures, indicating a Curie-Weiss like behavior, which is naturally ascribed to a local moment effect. Here, we noticed that the Fe's possibly occupy two different sites in Fe$_{1+\delta}$Te$_{1-x}$Se$_x$, i. e., Fe(I) occupies (0, 0, 0) site and has 1.6 $\sim$1.8 $\mu _B$ with itinerant characters and Fe(II) occupies (0.5, 0, z) with a localized moment of 2.5$\mu_B$. \cite{Bao,Chiba, Zhang} The localized moment has a strong competition with superconductivity, making the superconductivity very sensitive to the excess amount of Fe in the FeSe compound. \cite{Cava} Supposed by the band theory, the excess Fe occurs as Fe$^+ $, donating one electron to the Fe(I) layer. Experimentally, there is always excess iron in Fe(II) site and the number is lager in FeTe than that in FeSe, \cite{Sales} which may be the reason for that FeSe has such a high $T_c$ under high pressure while FeTe just changes to a metallic state at low temperatures in the same situation. \cite{Okada} In contrast to the doped sample, the end parent compound Fe$_{1+\delta}$Se shows a very weak $T$-linear behavior in $\chi(T)$, in agreement with the $^{77}$Se-nuclear magnetic resonance(NMR) measurement, \cite{Kotegawa} confirming a good reliability of Honda-Owen method. In NMR measurement, the Fe(I) site contributes completely to the Knight shift, strongly indicating that the Fe(I) site plays the key role to understand the superconducting mechanism in this system. \cite{Kotegawa} Furthermore, density functional calculations show that the electronic states near the Fermi level are mostly of Fe $3d$ characters from the Fe(I) site and with a smaller contribution from the excess Fe(II) site. \cite{Zhang} Herein, we ascribe the temperature dependence of the magnetic susceptibility in Fe$_{1+\delta}$Te$_{1-x}$Se$_x$ primarily originates from the Fe(I) and Fe(II) sites. However, the magnetic susceptibility of Fe(II) will be dominant since it has a larger local moment than that of Fe(I) site, especially at low temperatures. We suggest that the upturn in $\chi (T)$ at low temperatures comes from the excess Fe(II) site. In order to separate the contributions from the two different Fe sites  to the magnetic susceptibility, We fitted the magnetic susceptibility data with the Curie-Weiss law at low temperatures for sample with $x$ = 0.12, 0.20, 0.28 and 0.33 in the temperature range of 20 K $\leqslant$ T $\leqslant$ 50 K as
\begin{equation}
\chi(T)= \chi_0 +\frac{C}{(T-\theta)},                                
\end{equation}
where the $T$-independent term $\chi_{0}$ contains the Pauli paramagnetic susceptibility from itinerant-electron bands, the Van Vleck-orbital susceptibility and the Larmor diamagnetic susceptibility from ionic cores, $C$ stands for the Curie constant, and $\theta$ the Weiss temperature. Here, the Curie-Weiss term may be due to the Fe(II) site contribution. Therefore, the magnetic susceptibility of the Fe(I) can be roughly estimated as $\chi_{Fe(I)}$ = $\chi(T)$ - $C_{II}\over (T-\theta_{II})$, as shown in Fig. 2 (c), where $C_{II}$ is Curie constant due to the Fe(II) site, $\theta_{II}$ the Weiss temperature due to the Fe(II) site. We also fitted the data with Eq. (1) by using different temperature range of 100 K $\leqslant$ T $\leqslant$ 300 K for samples with $x$ = 0 and 0.05 (the nominal composition is $x$ = 0 and 0.10). The fitting results are listed in Table I in detail.
\begin{table}[htdp]
\caption{Fitted parameters using Eq. (1) for Fe$_{1+\delta}$Te$_{1-x}$Se$_{x}$ system as well as the real compositions checked by EDXS. $C$ and $\theta$ are obtained from the wider temperature fitting. The units of $C$, $\theta$, $\mu_{eff}$, C$_{II}$ and $\theta_{II}$ are emu K/mol, K, $\mu_B$, emu K/mol and K, respectively.}
\begin{center}
\begin{tabular}{lcccccccccc}
\hline\hline
sample(nominal) & Fe& Te& Se&$C$ & $\theta$ & $\mu_{eff}$ &C$_{(II)}$ & $\theta_{II}$  \\ \hline
  0 &1.12& 1 & 0& 2.24 & -319 & 4.2 &  -&-  \\  
 0.10&1.00  &0.95&0.05&1.6  & -260  & 3.7 &-  & -  \\ 
0.20 & 1.01 &0.88  & 0.12&- &-&-&0.10&  -52\\ 
0.30 &1.07  & 0.80 &0.20& -&-&-&0.02 & -5  \\ 
0.40(I)&1.12&0.72& 0.28&-&-&&0.12 & -24\\
0.40(II)&1.04& 0.67 & 0.33&-&-&-&0.05&-23  \\ 
 \hline\hline
\end{tabular}
\end{center}
\label{default}
\end{table}

 After subtracting the Fe(II) contribution from the susceptibility, it is clear that $\chi_{Fe(I)}$ decreases gradually from 300 K down to 20 K, as shown in Fig. 2 (c), qualitatively consistent with our NMR results. \cite{Our} It is important to note that there are other systems which also show the linear-in-$T$ behavior: for example, the geometric frustrated system Na$_{0.5}$CoO$_2$, \cite{Maw} the  high temperature cuprate superconductor La$_{2-x}$Sr$_x$CuO$_2$, \cite{Takagi} the simply metal Cr as well as its alloys \cite{Fawcett} and even the new discovered Fe-As based superconductors. \cite{Klingeler} All the above systems share a common feature: having antiferromagnetic spin fluctuations on their backgrounds. Very recently, Han and his colleagues studied the electronic structure and magnetic interaction in Fe$_{1+\delta}$Te. They found that the small amount of excess Fe played an important role in determining the magnetic structure and drove the Fermi surface nesting from ($\pi$, $\pi$) to ($\pi $, 0). \cite{Han} With increasing Se doping, the ratio of Fe(II) was depressed in Fe$_{1+\delta}$Te$_{1-x}$Se$_x$ system. \cite{Sales} Thus, the upturn at low temperatures will disappear in Se rich sample. It did so in the samples with $x$ = 0.45 and 0.48 as shown in Fig. 2 (c) in which we did not subtract the Fe(II) site contribution to the magnetic susceptibility but only show the raw data. In LaFeAsO$_{1-x}$F$_x$, the linear-in-$T$ behavior is considered to be a strong AF spin fluctuations with multi-orbital character. \cite{Klingeler} Korshunov argued that it was universal for systems with the strong ($\pi$, $\pi$) SDW fluctuation. \cite{Korshunov} In fact, we observed the $\mathbf{q} \ne 0$ modes of antiferromagnetic spin fluctuations were strongly enhanced toward $T_c$ in the normal state. \cite{Our} Overall the linear-in-$T$ behavior of $\chi(T)$ observed in our  single crystal samples strongly supports the above model, suggesting the importance of the ($\pi$, $\pi$ ) AF spin fluctuations originated from the Fe(I) site in superconducting mechanism of this system. 
 
\subsection{Superconducting State and Upper Critical Field $\mu_0H_{c2}^{orb}$}

The superconducting transition temperature was found to be $\sim$10 K for the sample with $x>$ 0.05 as shown in Figs. 3 (a), (b), (c), (d) and (e). Because the excess Fe in the Fe(II) site has a localized moment, \cite{Chiba, Zhang, Liu} where there is the more excess amount of Fe(II), the less superconducting volume fraction is observed, compared with Figs. 3 (d) and (e), where these two samples have very close composition. With increasing the Se content $x$, the superconducting volume fraction was enhanced greatly. The susceptibility measured in the FC process shows no negative sign but a small positive value in the superconducting state, indicating an intrinsic pinning effect in this layered structure compound. \cite{Iye}
\begin{figure}[htbp] 
  \centering
 \includegraphics[width= 5.5cm]{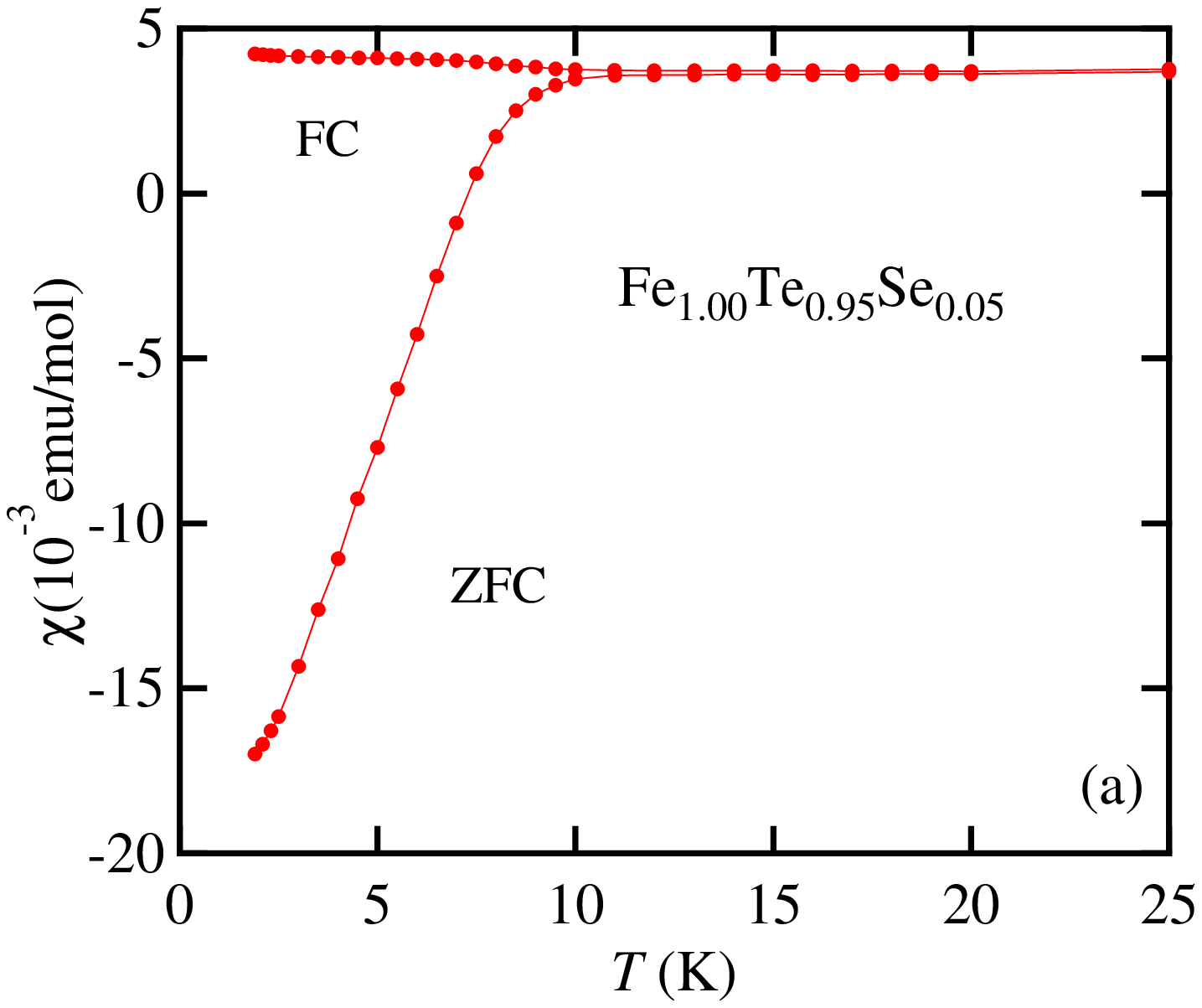}
  \includegraphics[width= 5.5cm]{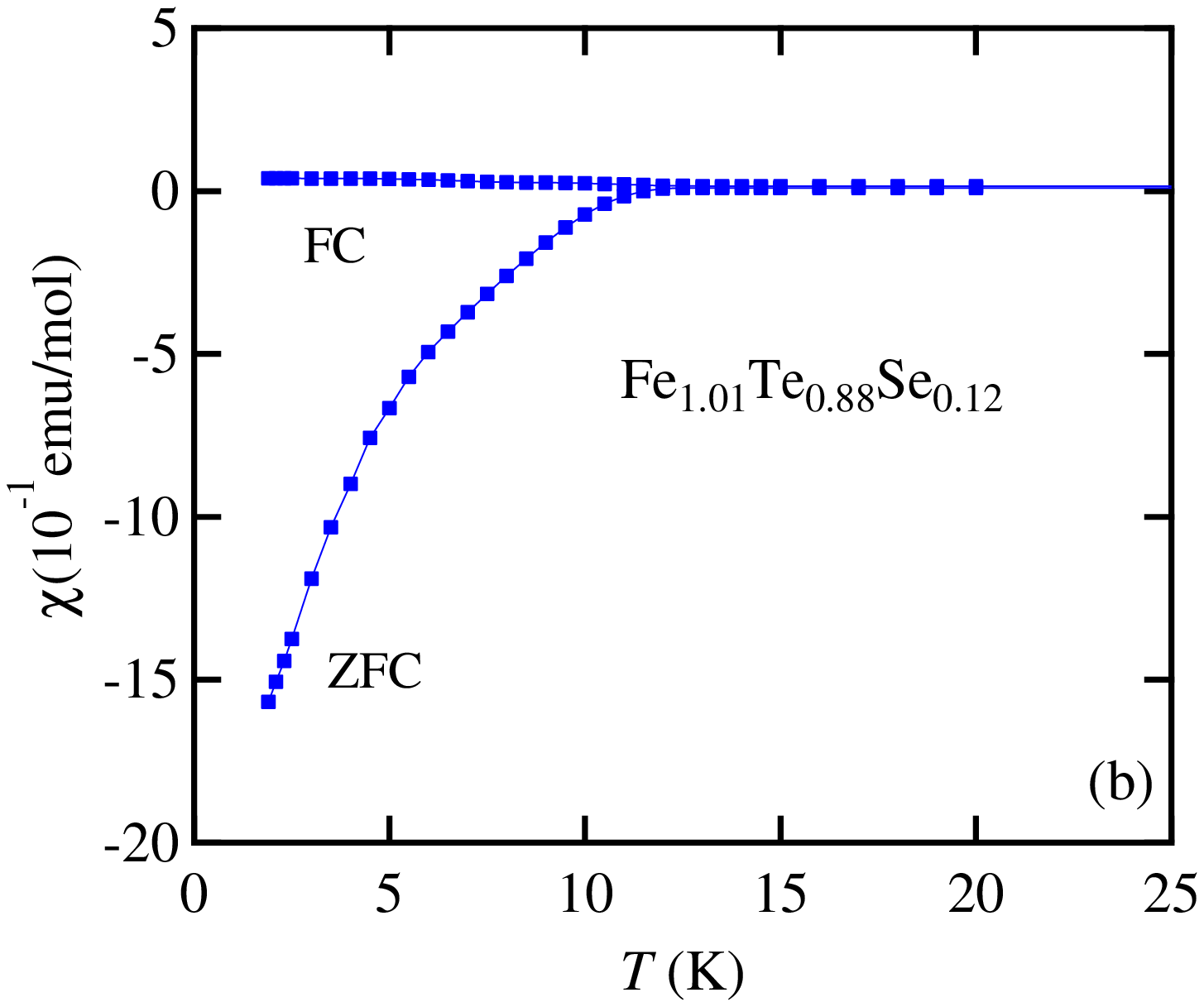}
   \includegraphics[width= 5.5cm]{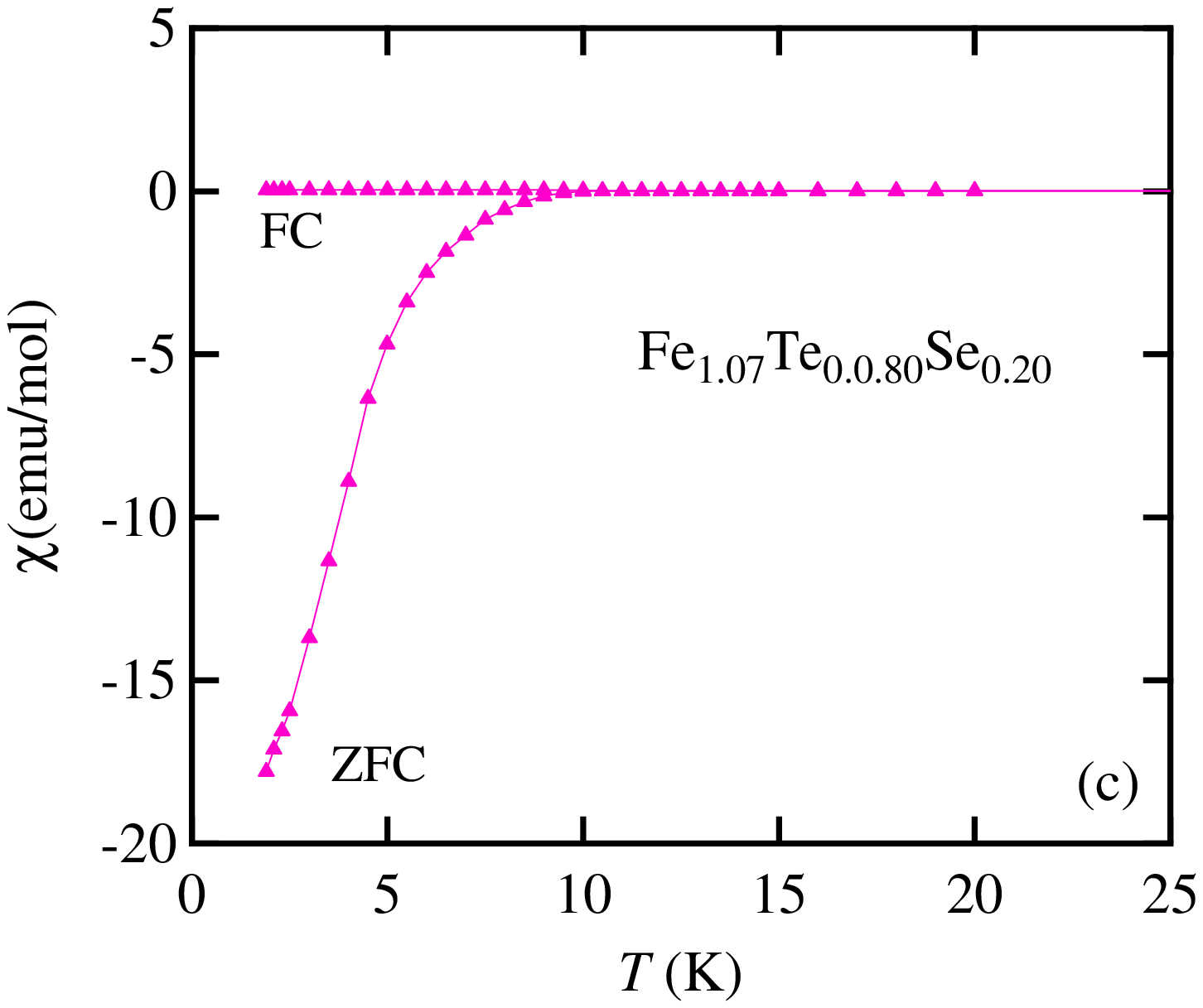}
    \includegraphics[width= 5.5cm]{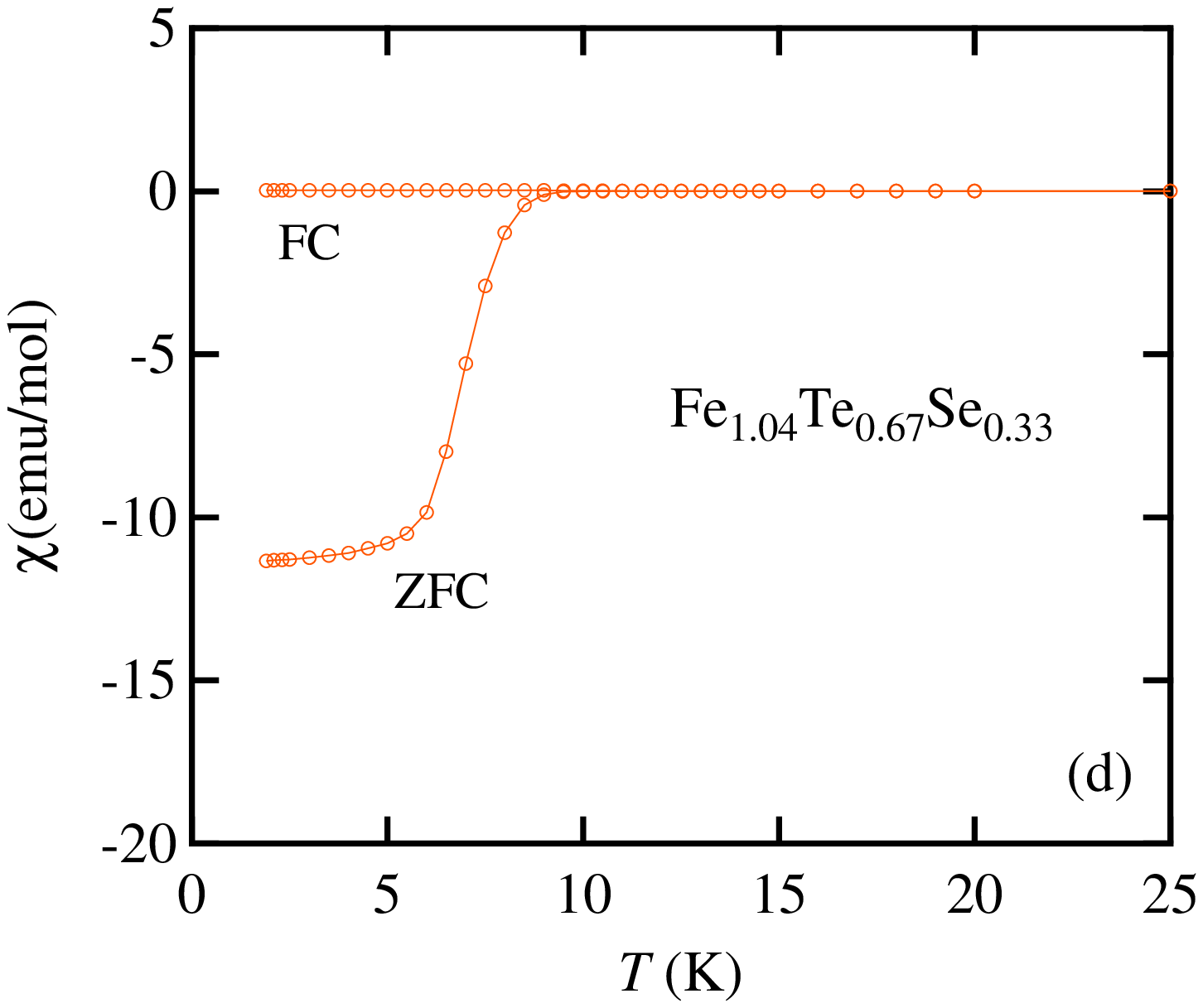}
     \includegraphics[width= 5.5cm]{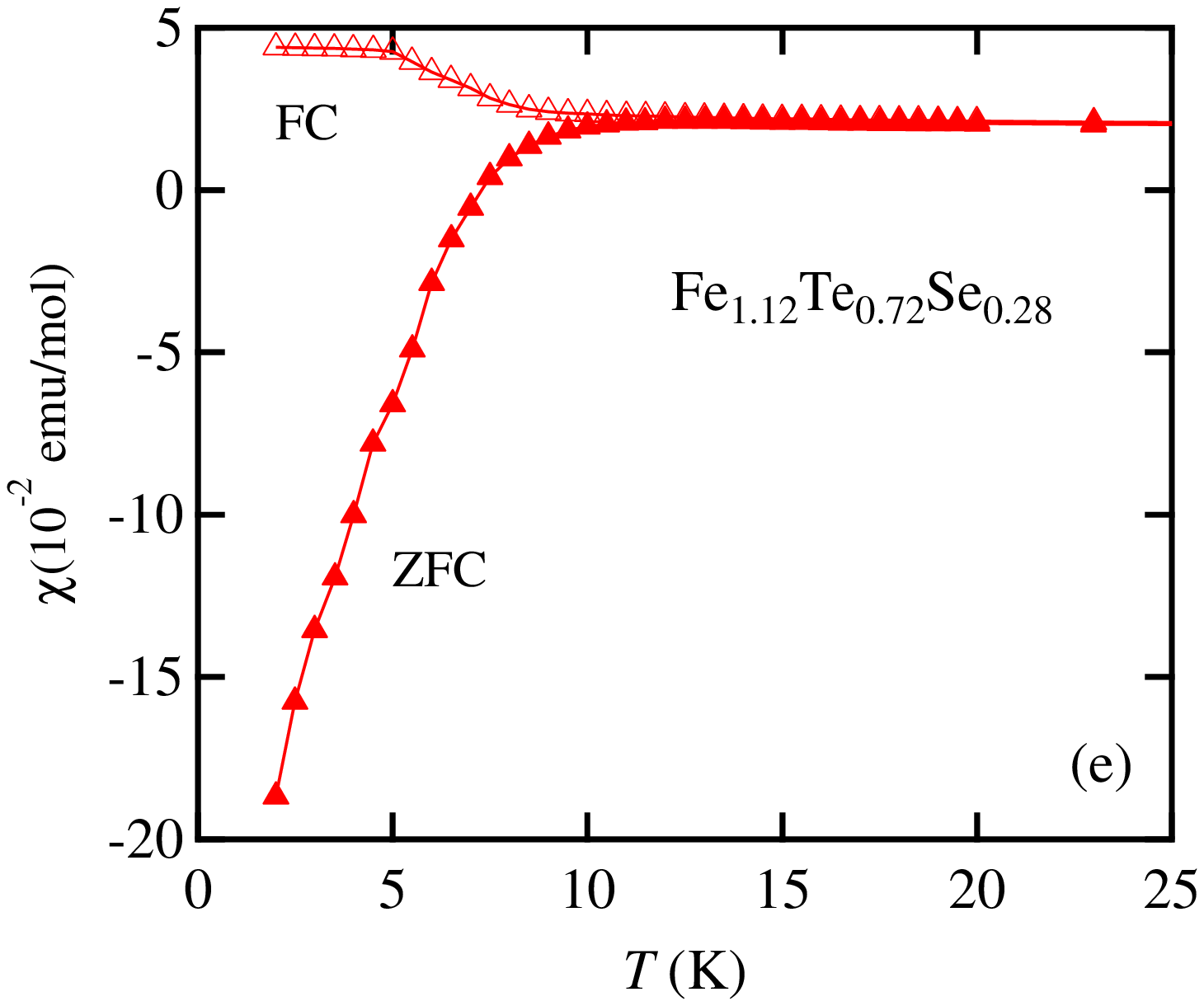}
  \caption{Temperature dependence of susceptibility in superconducting samples under magnetic field $H$ = 20 Oe applied along the c-axis $H$//c, for ZFC and FC processes. (a) Fe$_{1.00}$Te$_{0.95}$Se$_{0.05}$. (b) Fe$_{1.01}$Te$_{0.88}$Se$_{0.12}$, (c) Fe$_{1.07}$Te$_{0.80}$Se$_{0.20}$. (d) Fe$_{1.04}$Te$_{0.67}$Se$_{0.33}$. (e) Fe$_{1.12}$Te$_{0.72}$Se$_{0.28}$.}
  \end{figure}
  
In order to estimate the superconducting parameters, we selected two samples with close composition Fe$_{1.12}$Te$_{0.72}$Se$_{0.28}$ (simplified as R1) and Fe$_{1.04}$Te$_{0.67}$Se$_{0.33}$ (simplified as R2) for resistivity measurements. Figures 4 (a), (b), (c) and (d) show the suppression of the superconducting transition in the   electrical resistivity for $H$//c and $H$//a up to 14 T for the samples of R1 and R2, respectively. With increasing the magnetic field, the superconducting transitions shifted to lower temperatures, and  became broadened. The upper critical field $\mu_0H_{c2}^{orb}$ determined from the onset $T_c$ were plotted in Figs. 5 (a) and (b) for the samples R1 and R2, respectively. Here, the onset $T_c$ was defined as the resistivity falls to 90\% of the $\rho_0$ value in the normal state  just above $T_c$. The initial slopes $\partial \mu_0 H_{c2}/\partial T $ near $T_c$ are -6 T/K and -3.9 T/K for the R1 sample with $H$//a and $H$//c, respectively, leading to an estimation of the upper critical field extrapolated to zero-temperature as $\mu_0H_{c2}^{orb}(0)$ = 57 T and  37 T, by using the Werthamer-Helfand-Hohenberg (WHH) model as 
  \begin{equation}
  \mu_0 H_{C2}^{orb}(0) = -0.693 T_c (\frac{\partial \mu_0H_{C2}^{orb}}{\partial T})_{T=T_c}.
  \end{equation}
  In contrast to the sample R1, the sample R2 shows larger initial slopes of -8.7 T/K and -4.2 T/K as well as the upper critical fields of 85 T and 40 T under $H$//a and $H$//c, respectively. The upper critical field in this system is comparable with the cases of the Fe-As based superconductors LaFeAsO$_{0.93}$F$_{0.03}$ \cite{Kohama} and KFe$_2$As$_2$. \cite{Terashima} In addition, the upper critical field is much lager than the Pauli limit field $\mu_{0}H_{\mathrm{P}}$ =1.84 $T_{c}$ $\sim$ 25 T. The anisotropy coefficients $\Gamma$ determined from $\Gamma$ = $H_{c2}^{\perp}$/$H_{c2}^{\parallel}$ are 1.54 and 2.1 for the samples R1 and R2, respectively. In fact, the $\Gamma$ is rather isotropic at low temperatures, indicating the three dimensional nature of the Fermi-surface topology.\cite {Yuan} These results strongly suggest an unconventional superconducting mechanism in this compound. The estimated parameters are listed in Table II.
  \begin{figure}[htbp] 
  \centering
 \includegraphics[width= 6cm]{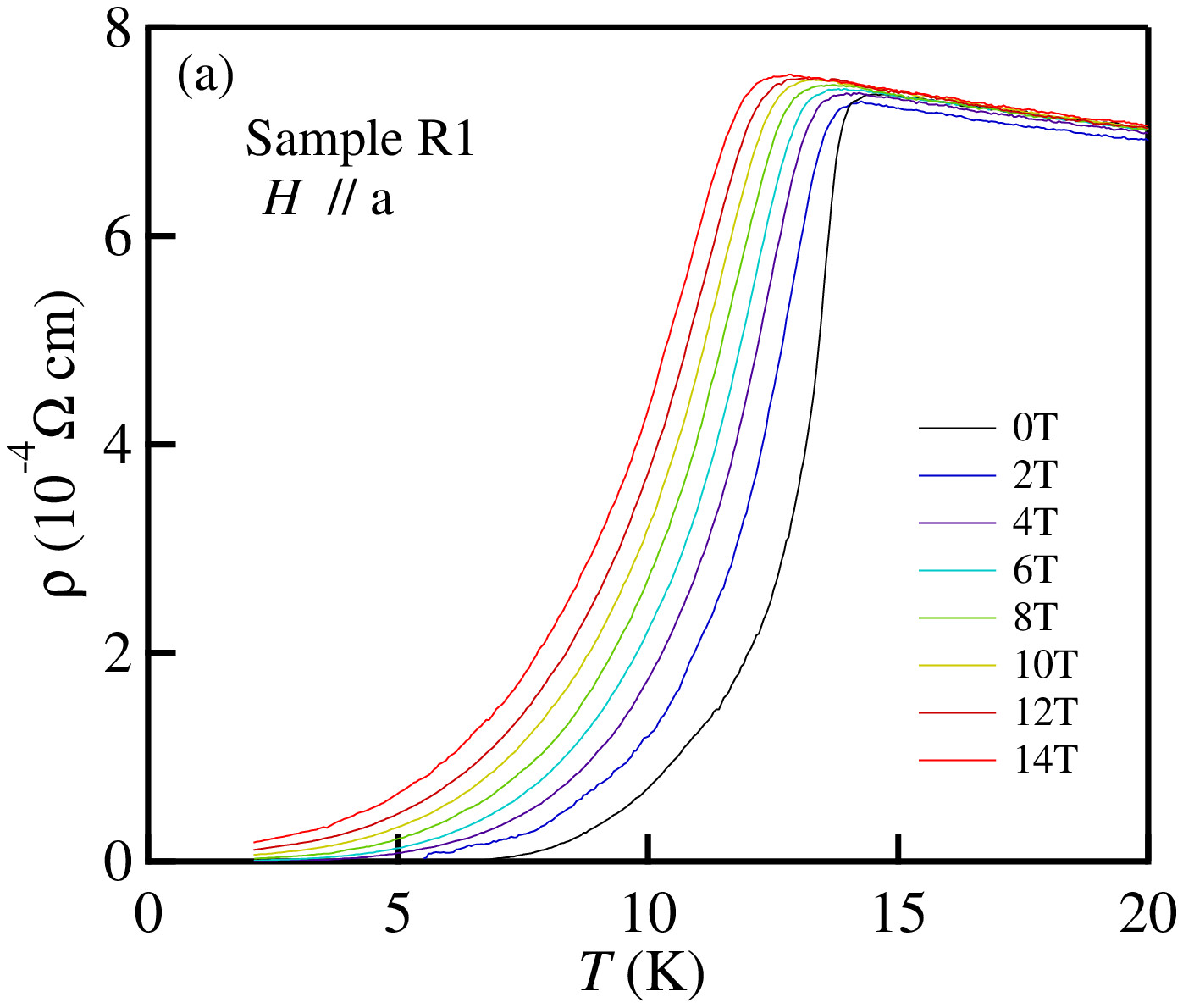}
  \includegraphics[width=6cm]{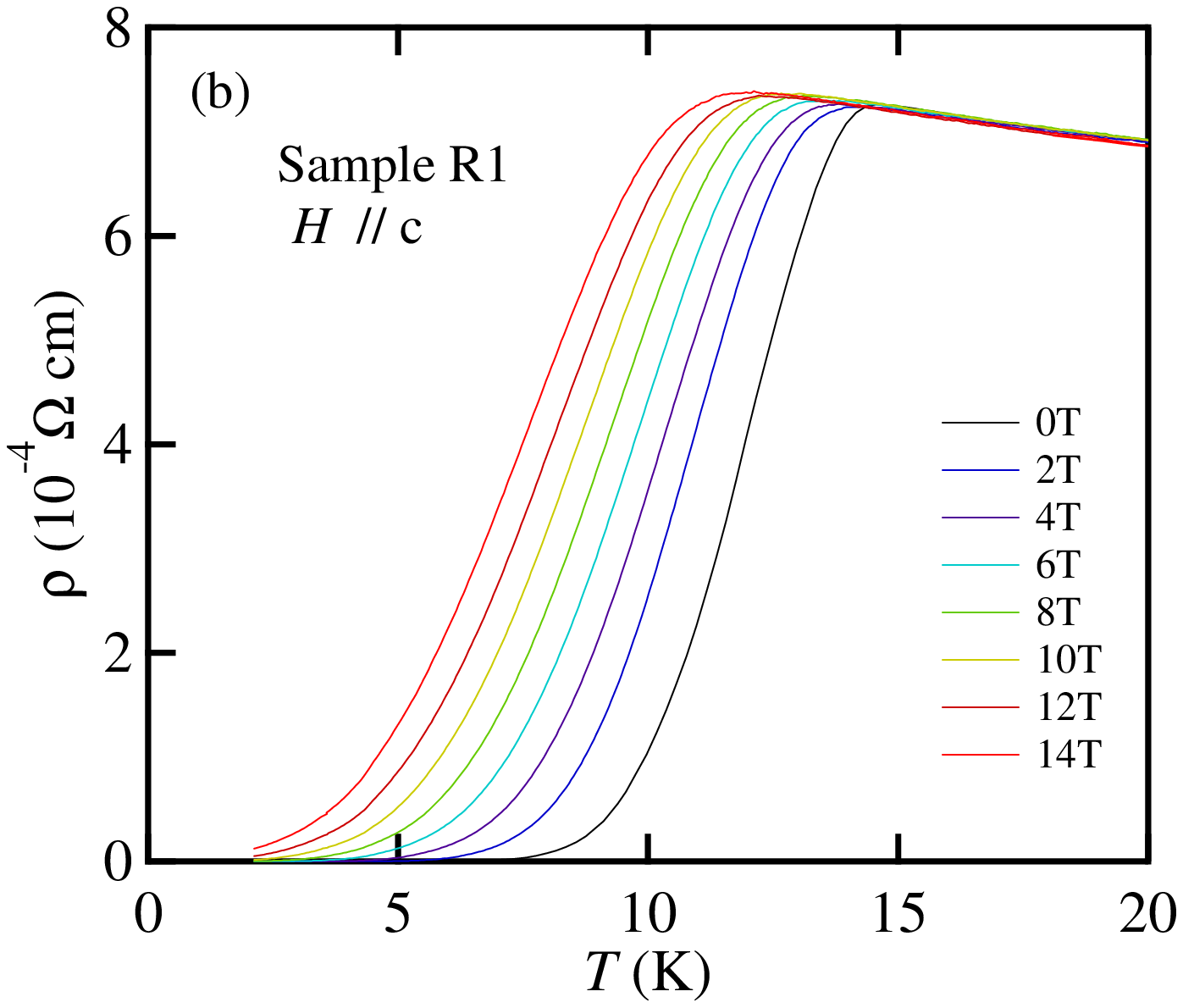}
 \includegraphics[width= 6cm]{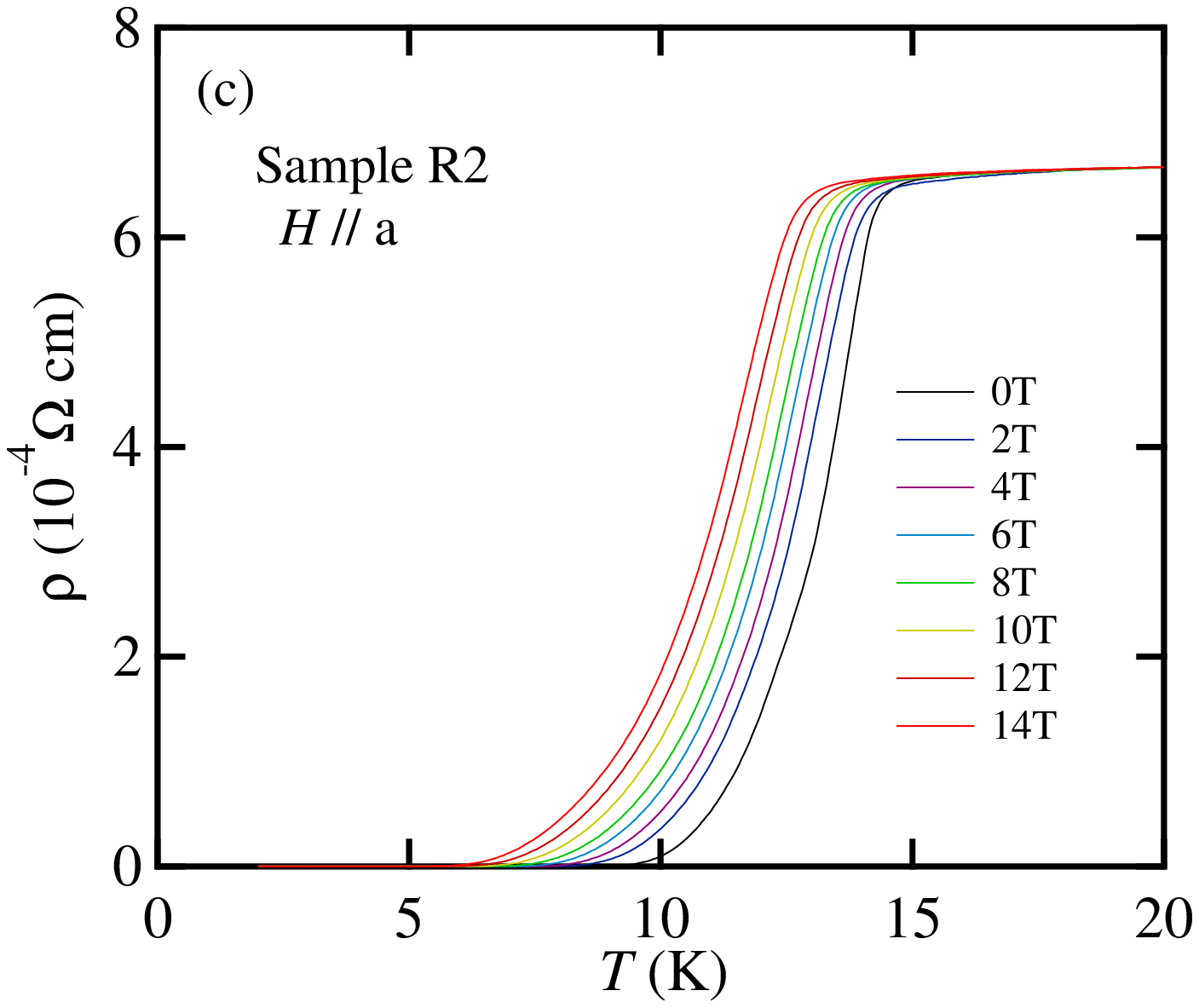}
 \includegraphics[width= 6cm]{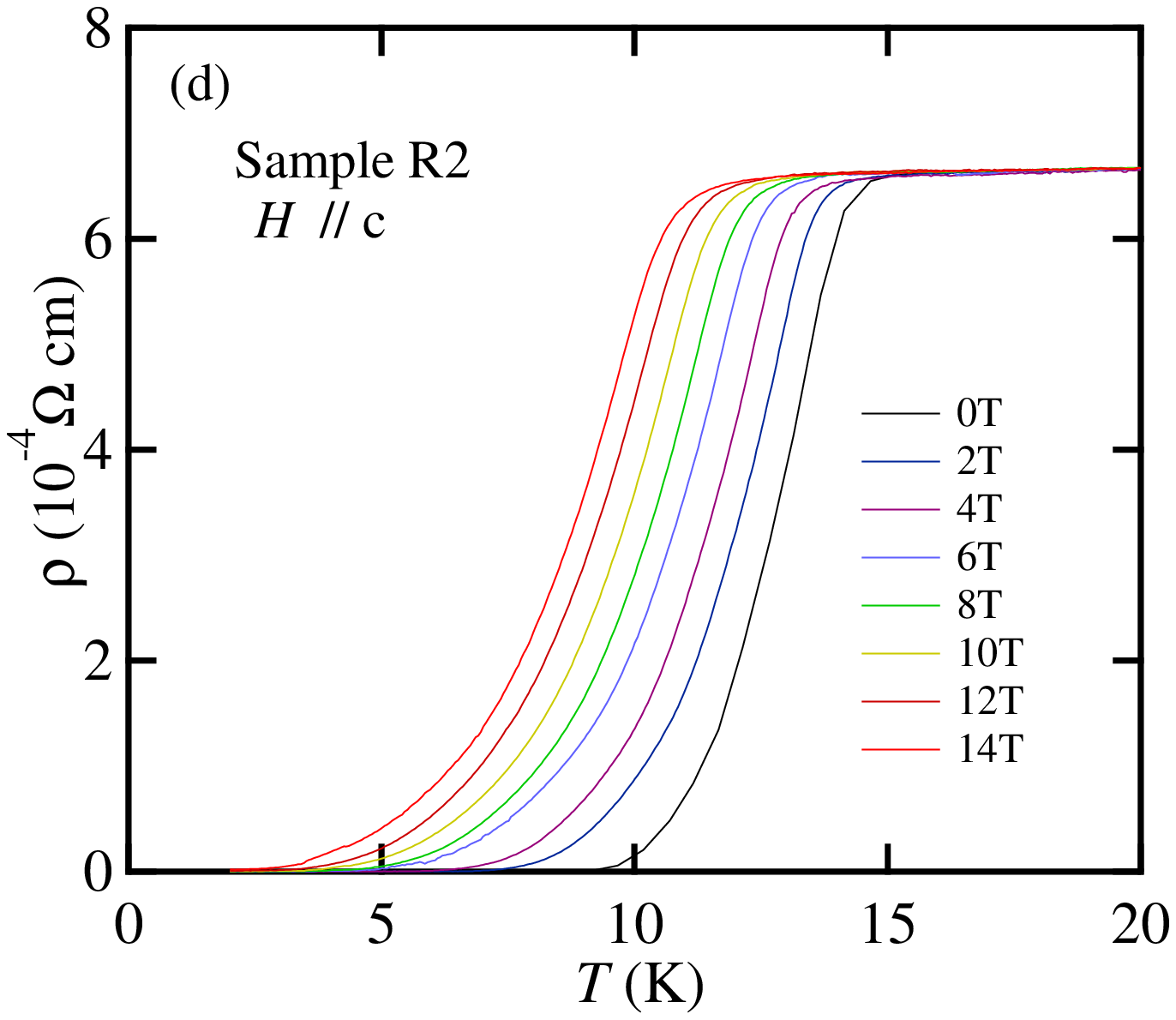}

 \caption{Temperature dependence of resistivity ($\rho$) under magnetic field ($H$) up to 14 T (0, 2, 4, 6, 8, 10, 12 and 14 T) for samples of R1 and R2: (a) The sample R1 for $H$ // c. (b) The sample R1 for $H$ // a. (c) The sample R2 for $H$ // c. (d) The sample R2 for $H$ // a.}
\end{figure}
  \begin{figure}[htbp] 
   \centering
 \includegraphics[width= 8cm]{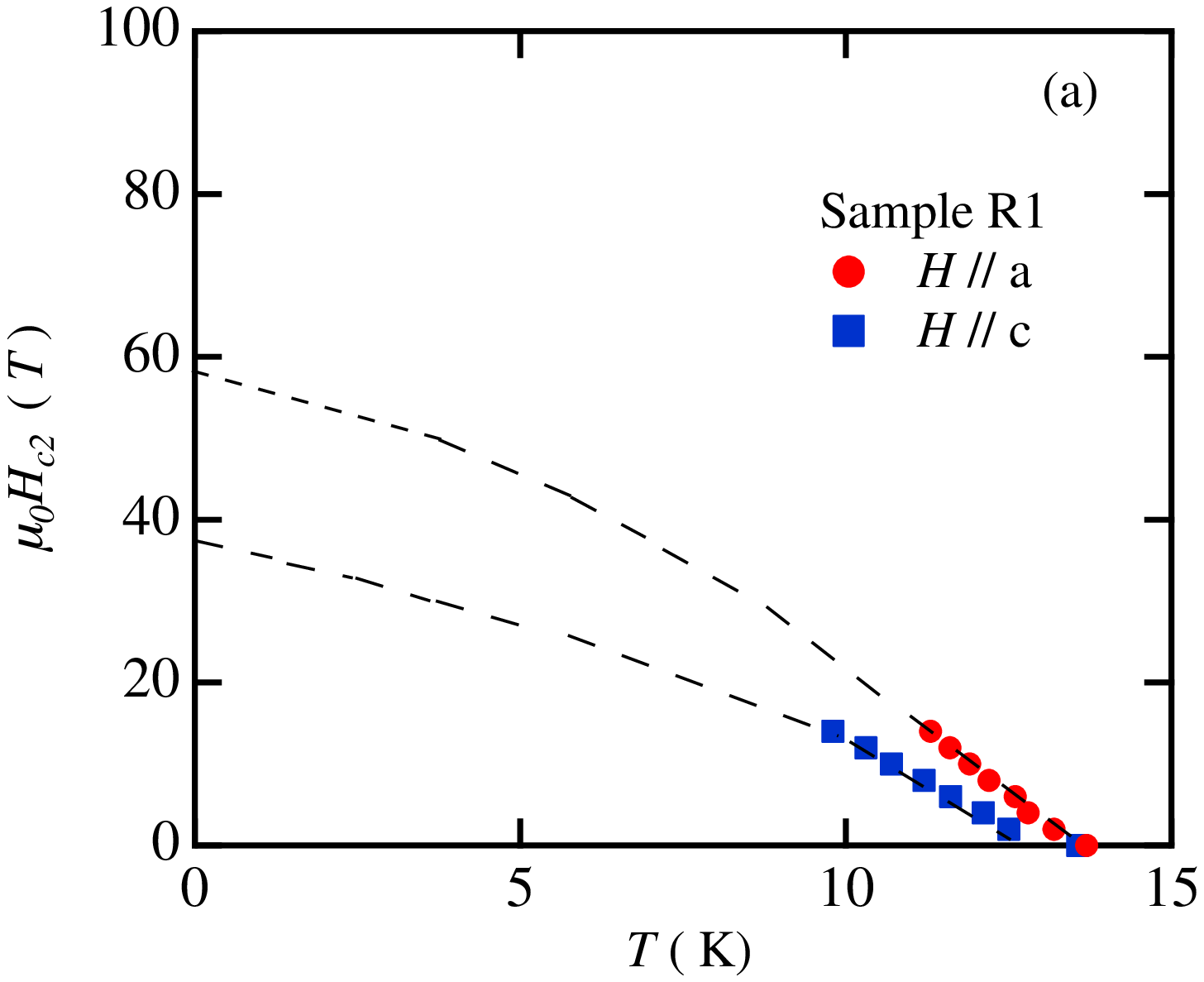}
  \includegraphics[width=8cm]{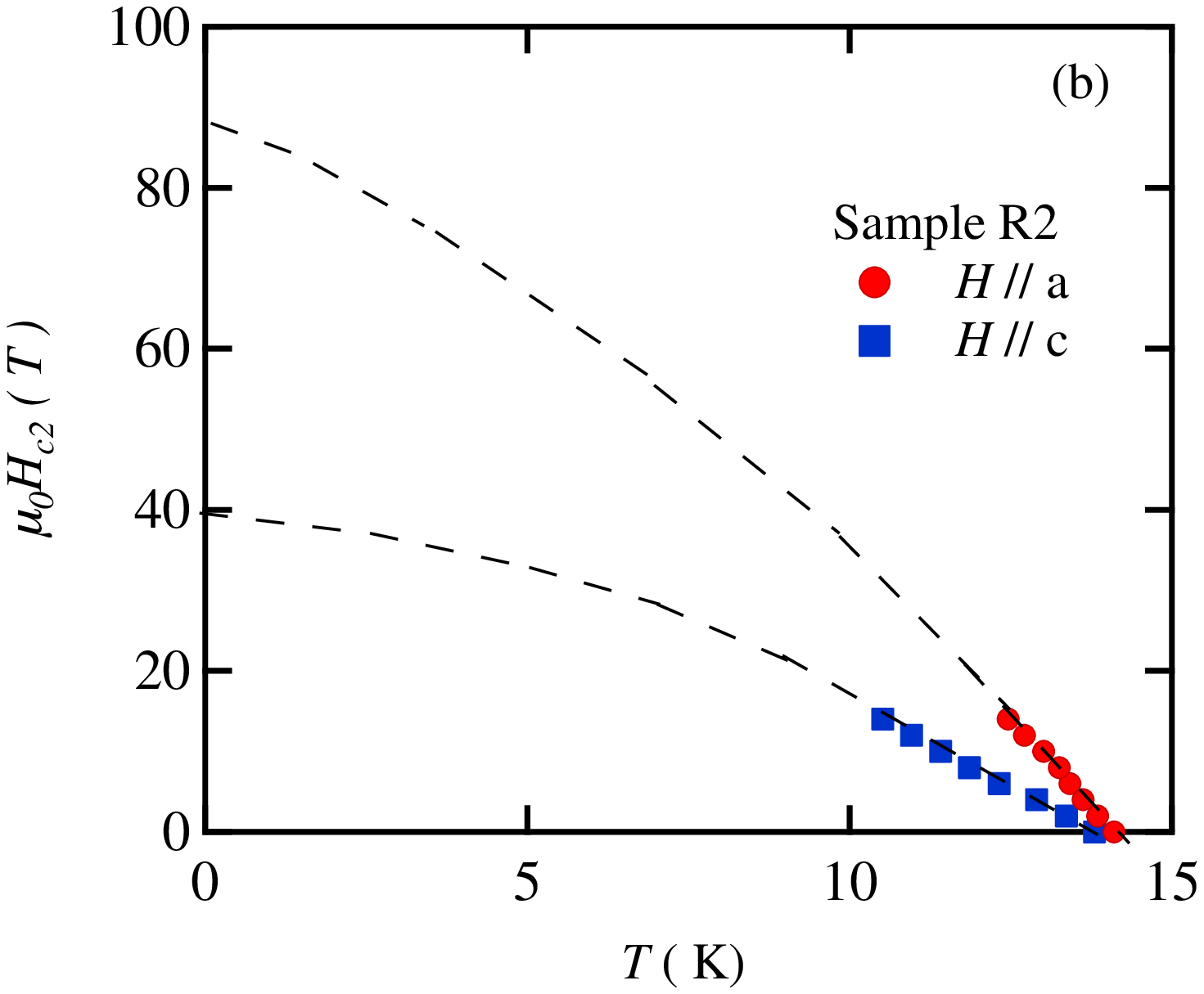}

 \caption{Temperature dependence of the upper critical fields of (a) the sample R1 and (b) the sample R2. The dashed line is the estimation by the WHH theory.}
\end{figure}
 
   The initial slope of $\mu_0H_{c2}^{orb}$ near $T_c$ is also weakly anisotropic. The similar behavior was also observed in the 122-type compounds, and was thought to be two-band superconductivity. \cite{Baily} The Sample R2 with the less Fe shows a larger initial slope and a higher upper critical field in compared with the sample R1, indicating the existence of Fe(II) affects the electronic band structure and consequently the superconductivity greatly. The initial slope near $T_c$ is proportional to the square of the electron effective mass $m^{*2}$, in agreement with a large $ \gamma \sim 39$ mJ/mol K$^2$. \cite{Sales} In strongly correlated electron systems, the Kadowaki-Woods ratio $A/\gamma^2$ is expected to be a constant $\sim$ 1.0 $\times$ 10$^{-5}$$ \mu\Omega$ cm (mJ/mol-K)$^2$, where $A$ is the quadratic term of the resistivity and $\gamma$ is the linear term coefficient of the specific heat, so called the electronic specific heat coefficient. We obtained $A \sim 0.03$ $\mu \Omega$ cm /K$^{2}$ by fitting the data with $\rho = \rho_0 +AT^2$ in the temperature range of 16 K $\leqslant$ T $\leqslant$ 20 K for the sample R2, resulting in $A/\gamma^2 \sim$ 2$\times$ 10$^{-5}$$\mu \Omega$ cm/(mJ/mol K)$^2$, which is a little bit larger than the value of heavy fermion compound UBe$_{13}. $\cite{UBe13} The Wilson ratio $R_w = \pi^2k_B^2 \chi_{spin}/3\mu_B^2 \gamma$ is estimated as 5.7 for the sample R2 with $\chi_{spin}$ of $2\times 10^{-3}$emu/mol from our data, well exceeding the unity for a free electron system. These results strongly suggest that the electron in superconducting Fe$_{1+\delta}$Te$_{1-x}$Se$_x$ is strongly correlated, being in good agreement with our recent NMR investigation on the same single crystal of Fe$_{1.04}$Te$_{0.67}$Se$_{0.33}$ which strongly indicates the unconventional d-wave superconductivity with spin singlet pairing-symmetry. \cite{Our} 
 
 It is also very important to know whether the Fe$_{1+\delta}$Te$_{1-x}$Se$_x$ is a clean superconductor or not. To solve this issue, we need to know the mean free path $\ell$ and the Pippard coherence length $ \xi_0$. On the basis of the Bardeen-Cooper-Schrieffer (BCS) theory and the Drude model, $\ell$ = $\hbar$ (3$\pi^2$)$^{1/3}$/e$^2$$\rho_0$n$^{2/3}$ and $\xi_{0}$ = $\hbar$V$_F$/$\pi$$\Delta$, where $n$ is the carrier concentration, $\rho_0$ the residual resistivity, V$_F$ the Fermi velocity and $\Delta$ the superconducting gap. Giving the superconducting gap 2$\Delta$/k$_B$$T_c$ = 3.52 in the BCS theory, the $\xi_0$ can be written as $\hbar^2$(3$\pi^{2}$n)$^{1/3}$/1.76$\pi$$m$$k_{B}$$T_{c}$, where $m$ is the free electron rest mass. Very recently, the angle-resolved photoemission spectroscopy (ARPES) measurement on Fe$_{1.03}$Se$_{0.3}$Te$_{0.7}$ showed the Fermi velocity $\sim$ 0.4 eV$\mathrm {\AA}$ for both the hole and the electron bands and the superconducting gap $\Delta$ $\sim$ 4 meV. \cite{Nakayama} Therefore the $\xi_0$ is estimated as 33.5$\mathrm {\AA}$ and the carrier concentration estimated as $\sim$ 6.8$\times$10$^{23}$/m$^3$. Since the composition is very close among the samples of Fe$_{1.12}$Te$_{0.72}$Se$_{0.28}$, Fe$_{1.04}$Te$_{0.67}$Se$_{0.33}$ and Fe$_{1.03}$Te$_{0.70}$Te$_{0.30}$ (in ref  \cite{Nakayama}), it is reasonable to consider that the carrier concentration does not change very much among these three samples. The residual resistivity was estimated as $\rho_0 = 0.70\times 10^{-5}\Omega m$, $0.65\times 10^{-5}\Omega m$ for the samples R1 and R2, respectively. Therefore, $\xi_0$ is estimated as $\sim$31 $\mathrm {\AA}$ and $\ell$ $\sim$ 2336 $\mathrm{\AA}$ for the sample R1 and $\xi_0$ $\sim$ 32 $\mathrm {\AA}$, $\ell$$\sim$ 2516 $\mathrm{\AA}$ for the case of the sample R2. The estimated parameters are listed in Table II. Therefore, the ratio of $\ell$/$\xi_0$ $\sim$ 80 is well exceeding the unity so that the Fe$_{1+\delta}$Te$_{1-x}$Se$_x$ is thought to be a clean superconductor and the estimated superconducting parameters are considered to be intrinsic. For example, in a clean superconductor, $\xi_{GL}$ $\sim$ 0.74 $\xi_0$/(1-$T/T_c$)$^{1/2}$. We estimated $\xi_{GL}$$\sim$ 24 $\mathrm{\AA}$ for sample R2 at T = 0 K along the c direction, in good agreement with the value of 29 $\mathrm{\AA}$ derived from the upper critical field, where $\xi_{GL}$ is expressed as $\xi_{GL}^2$ = $\phi_{0}$/2$\pi$$H_{c2}^{orb}$ ( $\phi_0$ = 2 $\times$ 10$^{-7}$ Oe cm$^2$). However, it should be pointed out that those estimations based on the one band theory which may not valid for the multi-band compound, and much information will be needed for further discussion on superconductivity in this system. 
 \begin{table}[t]
\caption{Estimated superconducting parameters for the samples R1 and R2. T$_c$ = 13.7 and 14.1 K for the samples R1 and R2, respectively.}
\begin{center}
\begin{tabular}{cccccccccc}
\hline\hline
sample& $\partial \mu_0H^{orb}_{c2}$/$\partial$T & $\mu_0$$H^{orb}_{c2}$ &$\mu_{0}H_{\mathrm{P}}$(0)&$\xi$$_{0}$&$\xi$$_{GL}$&$\ell$ &$\ell$/$\xi_0$\\ 
($H$//a,c)&($T/K$)&(T)&(T)&(\AA)&(\AA)&(\AA)&\\ \hline

R1(a) &-6.0&57&25&32&24&2336&73 \\ 
R1(c)&-3.9&37&25&32&30\\
R2(a)&-8.7&85&26&31&20&2516&79 \\
R2(c)&-4.2&40&29&32&29\\ \hline\hline
\end{tabular}
\end{center}
\label{default}
\end{table}
\section{CONCLUSIONS}
In summary, we successfully synthesized the single crystal of Fe$_{1+\delta}$Te$_{1-x}$Se$_x$ ($x$ = 0, 0.05, 0.12, 0.20, 0.28, 0.33, 0.45, 0.48 and 1.00; $0<\delta<0.12$) and measured their magnetic susceptibilities. The intrinsic magnetic susceptibility was obtained though Honda-Owen method for the first time. The nearly linear-in-$T$ behavior in susceptibility was observed in superconducting samples with $x$ = 0.12, 0.20, 0.28, 0.33 0.45, 0.48 and 1.00, indicating a close relationship between the AF spin fluctuations around ($\pi$, $\pi$) and the superconductivity. The excess Fe has a localized moment which affects the superconducting state greatly. The intrinsic susceptibility shows a weakly anisotropic behavior. Also, the initial slope near $T_c$ and the upper critical field estimated by measuring the resistivity under high magnetic fields show a weakly anisotropic behavior. The estimated coherence length $\xi_{GL}$, the Pippard coherence length $\xi_0$ and the mean free path $\ell$ by the BCS theory and the Drude model support a clean superconductor scenario. The estimations of Kadowaki-Woods and Wilson ratios indicate Fe$_{1+\delta}$Te$_{1-x}$Se$_x$ belongs to a strongly electron-correlated system. Consequently, the superconductivity in Fe$_{1+\delta}$Te$_{1-x}$Se$_x$ is considered to be of unconventional and in the strongly correlated one with the very high value of $\mu_0H_{c2}^{orb}$, which are also supported by our recent NMR study. \cite{Our}
\section*{Acknowledgement}
This research was supported by Grant-in-Aid for the Global COE Program ``International Center for Integrated Research and Advanced Education in Materials Science" and for Scientific Research on Priority Area ``Invention of anomalous quantum materials"(16076210) from the Ministry of Education, Culture, Sports, Science and Technology of Japan, and also by Grants-in-Aid for Scientific Research (19350030) from the Japan Society for Promotion of Science.

\end{document}